\RequirePackage{lineno}
\documentclass[aps,prc,twocolumn,superscriptaddress, showpacs]{revtex4}

\usepackage{epsfig}
\usepackage{amssymb}
\usepackage{bm}
\usepackage{graphicx,color}
\usepackage{dcolumn}
\usepackage{longtable}
\usepackage{epstopdf}
\usepackage{csquotes}

\begin{document}



\title{Calculations of Neutron Reflectivity in the eV Energy Range from Mirrors made of Heavy Nuclei with Neutron-Nucleus Resonances}
\author{W. M. Snow}
\email[Corresponding author: ]{wsnow@indiana.edu}
\author{K. A. Dickerson}
\author{J. S. Devaney}
\affiliation{Indiana University/CEEM, 2401 Milo B. Sampson Lane, Bloomington, IN 47408, USA}
\author{C. Haddock}
\affiliation{Center for Neutron Research, National Institute for Standards and Technology, Gaithersburg, MD 20899, USA }

\date{\today}

\begin{abstract}
We evaluate the reflectivity of neutron mirrors composed of certain heavy nuclei which possess strong neutron-nucleus resonances in the eV energy range. We show that the reflectivity of such a mirror for some nuclei can in principle be high enough near energies corresponding to compound neutron-nucleus resonances to be of interest for certain scientific applications in non-destructive evaluation of subsurface material composition and in the theory of neutron optics beyond the kinematic limit.

\end{abstract}



\maketitle

\section{Introduction}

Neutron optics based on mirror reflection~\cite{Fermi46, Fermi47} from the neutron optical potential of matter, which is composed from the spatial average of the individual neutron-nucleus scattering amplitudes from the nuclei in the medium, has been widely used in slow neutron instrumentation now for decades. The ability to conduct measurements using slow neutrons far from the radiation backgrounds and noisy environment near the neutron source without $1/r^{2}$ intensity falloff revolutionized the field and greatly expanded its range of scientific applications~\cite{Leibnitz63, Leibnitz66}. Early slow neutron mirrors were made from nuclei with relatively large neutron optical potentials. $^{58}$Ni was (and still is in many cases) a common choice. 

Supermirror neutron guides~\cite{Turchin67, Mezei76, Mezei77} also employ the neutron optical potential to reflect neutrons. Neutron supermirrors are engineered to realize a reflectivity whose critical angle for near-total external reflection is larger than that from the neutron optical potential of a uniform medium. This is done using multilayer coatings of materials with a large contrast in the neutron scattering length density. Incident neutrons diffract from the one-dimensional crystal created by these stacked multilayers, and the constructive interference from this diffraction scattering produces a peak in the neutron reflectivity for specific values of the transverse momentum which match the diffraction condition~\cite{Hayter89, Schaerpf89}. If one deposits a set of layers with a continuous distribution of bilayer separations which cover the transverse momentum phase space in a neutron beam, one can transport all of the neutrons of interest. In a typical beam from a cold neutron source, one can use supermirrors to efficiently transport a much larger fraction of the beam phase space from the source over distances as long as 100 meters or more compared to a mirror made of a single material. Supermirror guides are typically characterized by a dimensionless number $m$ which is the ratio of the critical angle for near-total reflection from the guide to the same critical angle for a neutron mirror composed of natural nickel. In these units $^{58}$Ni corresponds to $m=1.2$. Supermirror guides with $m=7$ are now commercially available.   

eV neutron beams are used in some measurements in neutron scattering~\cite{Lovesey84, Carpenter2015} and neutron imaging for condensed matter and materials studies~\cite{Windsor81}. The most heavily used neutron scattering application for eV neutrons are measurements of the longitudinal momentum distribution of hydrogen, deuterium, and other light atoms in materials~\cite{Andreani05} and vibrational spectroscopy~\cite{Mitchell05}. Neutron resonance imaging~\cite{Behrens1979} to get nondestructive quantitative information on the local environment of specific isotopes embedded in materials from the shapes of the Doppler broadening of the resonance linewidths also of course operates in the eV neutron energy regime where resonances are abundant. With the development of a growing number of intense neutron sources in the eV energy range based on proton spallation~\cite{Carpenter77}, where the incompletely-moderated $1/E$ high energy tail of the neutron energy spectrum from a hydrogenous moderator is much richer in eV neutrons than is the high energy tail of the spectrum from a research reactor, it is well worth considering the possibility of developing other scientific applications of eV neutrons.


Both normal neutron mirrors and neutron supermirrors typically employ the real part of the (in general complex) neutron optical potential of the medium. This is because, in the meV energy range emitted by slow neutron sources, the neutron-nucleus scattering amplitude has a real part that is typically much larger than the imaginary part. This follows in turn from probability conservation as embodied in the optical theorem of nonrelativistic scattering theory, $Im[f(\theta=0)]={k \sigma \over 4\pi}$ where $k$ is the neutron wave vector, $\sigma$ is the total cross section, and $f(\theta=0)$ is the forward scattering amplitude. The slow neutron energy regime corresponds to $kR \approx 10^{-4}$ where $R$ is the potential scattering range from the neutron-nucleus interaction, which implies that the scattering is dominated by the s-wave component of the partial wave expansion of the scattering amplitude. For this case, $f(\theta=0)=f$ is of order $R$ in magnitude and $\sigma=4\pi |f|^{2}$, so the optical theorem implies $Im[f]=k[Re(f)^{2}+Im(f)^{2}] \approx kR^{2}\approx 10^{-4} |f|$.   

This argument breaks down however for inelastic scattering and for resonance scattering. For the case of resonance scattering of most interest in this work, $f$ can be much larger than $R$. The simplest case of elastic scattering with resonances will serve to make the point clear. One can decompose the partial wave expansion in nonrelativistic scattering theory into terms associated with ingoing and outgoing spherical waves~\cite{Sakurai}

\begin{equation}
    \label{eq:partialwave}
 f(\theta)=\sum_{l=0}^{\infty}{(2l+1)P_{l} \over 2ikr}[(1+2ikf_{l}(k)){e^{ikr}}-{e^{-i(kr-l\pi)}}]
\end{equation}

\noindent where $f_{l}(k)$ is the partial wave scattering amplitude in orbital angular momentum channel $l$ and $P_{l}(\cos{\theta})$ is the $l^{th}$ Legendre polynomial. As elastic scattering merely redirects the incident wave packet, the optical theorem demands that $|1+2ikf_{l}(k)|=1$. This condition can be written in terms of the elastic scattering phase shift $\delta_{l}$ as $1+2ikf_{l}(k)=\exp{2i\delta_{l}}$, which implies 

\begin{equation}
    \label{eq:phaseshift}
kf_{l}={{\exp{(2i\delta_{l})}-1} \over {2i}}={1 \over {\cot{\delta_{l}}-i}}\,\cdot
\end{equation} 

In resonance scattering near the resonance energy $E_{r}$ the phase shift increases rapidly from $0$ to $\pi$. On resonance $\delta_{l}=\pi/2$ and $\cot{\delta_{l}}=0$ so $f_{l}=i/k$ is purely imaginary and also takes the maximum possible value consistent with unitarity. In particular $f_{l}$ need not be of order $R$.   Expanding  $\cot{\delta_{l}}$ near the resonance energy as a power series in $E-E_{r}$ and keeping only the linear term, $\cot{\delta_{l}}(E)=0+(\Gamma/2)(E-E_{r})$  gives the well-known Breit-Wigner form 

\begin{equation}
    \label{eq:BreitWigner}
kf_{l}={{\Gamma/2}\over{E-E_{r}+i\Gamma/2}}\,\cdot 
\end{equation}

$f_{l}$ then both possesses a large imaginary component and also can be large compared to $R$ as long as $(E-E_{r}) \approx \Gamma$. 

Neutrons can coherently reflect from the imaginary part of the optical potential of the medium as well as from the real part. In this paper we evaluate the neutron reflectivity in the eV energy range of mirrors composed of heavy nuclei. The idea is to take advantage of the large imaginary component of the scattering amplitude present at and near neutron-nucleus resonance energies to enhance the reflectivity of neutrons. The large imaginary component of the scattering amplitude is only present for neutrons with energies within the width $\Gamma$ of the resonance energy, and since both $\Gamma/E_{r}\ll1$ and $\Gamma \ll \Delta E$ where $\Delta E$ is the typical separation between neighboring resonances for the great majority of low-lying n-A resonances, no single nucleus can be used to reflect a broad range of neutron energies using this mechanism. As the neutron optical potential for coherent reflection is simply a weighted linear sum of the amplitudes from all of the scatterers in the medium, one can imagine making a mirror out of a \lq\lq cocktail\rq\rq of different nuclei. Although as we will see below this idea does not seem practical, we will see that the reflectivity is large enough for certain nuclei at the resonance energies to be of interest for certain applications.  As long as the thickness of the mirror is great enough that one can neglect the transmission through the mirror for all of the relevant energies in the beam, one can apply the well-known formulae from the theory of neutron optics and use the very extensive set of measured data on n-A scattering resonances to calculate the reflectivity in an idealized limit. It is also important to take into account the fact that the neutron-nucleus resonances in the eV energy range typically posses a large inelastic component, usually dominated by (n, $\gamma$) reactions. Therefore the fraction of the coherently scattered neutrons on resonance is proportional to $\Gamma_{n} \over \Gamma_{tot}$ where $\Gamma_{n}$ and $\Gamma_{tot}$ are the neutron and total widths of the resonance, respectively. Typically ${\Gamma_{n} \over \Gamma_{tot}}\ll 1$, which further suppresses the reflection probability.

We were not able to find any previous considerations of this idea in the neutron optics literature, although they may well exist. In the case of ultracold neutrons with energies of hundreds of neV such calculations were actually done a very long time ago for the case of nuclei with large absorption cross sections near thershold~\cite{Gurevich1961} and it is well-known that ultracold neutrons bouncing from a mirror made of nuclei with a large neutron absorption cross section can possess a very large reflection probability. Frank and collaborators~\cite{Frank2003} analyzed the resonance component of the neutron optical potential of Gd for the interpretation of some ultracold neutron transmission experiments designed to investigate the $1/v$ neutron absorption law. By contrast our interest is in the eV neutron energy range. The effects of resonances in the theory of xray optics are by contrast well-developed in both theory and experiment~\cite{Attwood2000, Neilsen2001}.

The rest of this paper is organized as follows. In section 2 we review the relevant parts of the theory of neutron optics and of resonance scattering. We present and discuss the results of our calculations in section 3.  In section 4 we discuss some possible applications of our results. We conclude in section 5. 

\section{Neutron Optics Theory}
\label{sect:NeutOptTheory}
In this section we give a brief review of the relevant results
from the theory of neutron optics that are needed to understand
the foundation for the formulae used in this work. We also describe the regime of applicability of this theory and call attention to circumstances where it may break down. For a more
detailed treatment see Sears~\cite{Sea82}.

Neutron optics is based on the existence of the ``coherent wave''
which is the coordinate representation of the coherent state
formed  by the incident wave and the forward scattered wave in a
scattering  medium~\cite{Gol64}. It is determined by the solution of a one-body
Schrodinger equation

\begin{equation}
    \label{eq:coherent}
    \left[\frac{-\hbar^{2}}{ 2m}\Delta +v(r)\right]\psi(r)=E\psi(r)
\end{equation}
where $\psi(r)$ is the coherent wave and $v(r)$ is the optical
potential of the medium. The coherent wave satisfies the
Lippmann-Schwinger equation
\begin{equation}
    \label{eq:lippmann}
    \psi(r)=\left|k\right>+gv(r)\psi(r)
\end{equation}
where $\left|k\right>$ is the incident wave, $g$ is the one-body
Green's function for nonrelativistic motion  of a neutron and
$v(r)$ is the optical potential. The optical potential is related
to the one body $t$ matrix by
\begin{equation}
    \label{eq:tmatrix}
    t=v(r)+tgv(r)
\end{equation}
and this combination forms the usual coupled system of equations
of nonrelativistic scattering theory from a medium of a large number of scatterers. Given a form for the $t$
matrix  one can determine the optical potential and then solve the
one-body Schrodinger equation for the coherent wave.

One must make an approximation for the $t$ matrix of the neutron
in a medium of scatterers. The usual approximation is essentially
the Born approximation in which $v=t$ and
\begin{equation}
    \label{eq:tapprox}
    t=\sum_{l}t_{l}.
\end{equation}
Finally one must approximate the one-body $t$ matrix $t_{l}$.
Using the impulse approximation for scattering, one gets
\begin{equation}
    \label{eq:FermiApprox}
    t_{l}=~(2\pi \hbar^2/m)\sum_{l} b_{l} \delta({\bf r}-{\bf R}_{l}).
\end{equation}
Here, $l$ denotes the elemental species, $b_{l}$ is the coherent
scattering length for element $l$, ${\bf r}$ is a random spatial
coordinate and ${\bf R}_{l}$ defines the coordinate of each atom
the neutron can scatter from. From Eq.~\ref{eq:FermiApprox} we
then arrive at an expression for the optical potential

\begin{equation}
    \label{eq:opticalpotential1}
    v_{opt}(r)~=~(2\pi \hbar^2/m)\sum_{l} N_{l} b_{l},
\end{equation}

where $N_{l}$ is the number density of scatterers. In general the scattering amplitude, and therefore the optical potential,  is complex to account for
incoherent scattering,  absorption, and resonance contributions to the scattering amplitude. We refer to this result as the form of the optical potential in the kinematic limit.  

This relation between the neutron optical potential and the scattering amplitude is an approximation which neglects effects due to dispersion and multiple scattering in the medium. The approximation fails in two places:  in Eqs. \ref{eq:tapprox} and
\ref{eq:FermiApprox}, due to effects from atomic binding and multiple scattering respectively. Since the neutron-nucleus interaction is much stronger and much
shorter range than the binding forces of the atoms in matter, it
is reasonable to neglect the effects of chemical binding during
the neutron-nucleus collision. In addition, the short-range of the
interaction means that the timescale of the collision is much
shorter than the timescales associated with the motion of the atom
in the potential well. For both of these reasons, the $t$ operator
is usually approximated by the $t$ operator for a free atom. This
is known as the impulse approximation in scattering theory. In this approximation, the neutron optical properties of a medium
depend only on the coherent scattering length of the atoms and not
at all on the details of the binding of the atoms.  

Calculations of corrections to the impulse approximation for
neutron scattering lengths have been performed~\cite{Now82b,Die81,Now82a} which are consistent with the
optical theorem and reduce in appropriate limits to previous results
in the limit of static scatterers~\cite{Sea82, Lanz1997}. Intraparticle multiple scattering gives corrections of the same order of magnitude. In Eq.~\ref{eq:tapprox} the $t$ matrix of the
bound system of $N$ scatterers is expressed as the sum of the
(impulse approximation) one-body $t$ matrix. But this is known in
exact treatments of scattering theory to be an
approximation~\cite{Gol64}. The next order of approximation for the
$t$ matrix of the system is
\begin{equation}
    \label{eq:multiplescat}
    t= \sum_{l}t_{l} + \sum_{l,l^{'}, l\ne l^{'}} t_{l}Gt_{l^{'}}+\ldots
\end{equation}
where $G$ is the Green's function. To calculate the full $t$ operator of the system, which is what
is required to obtain the optical potential in
Eq.~\ref{eq:coherent} one must take multiple scattering also into
account. Then finally the optical potential, which is now a
function of the neutron momentum, must be solved from
Eq.~\ref{eq:tmatrix}.

Nowak~\cite{Now82b} performed this calculation to obtain the
modified expression for the index of refraction
\begin{equation}
    \label{eq:newindex}
    n^{2}=1-\frac{v(k)}{E}
\end{equation}
and two correction terms at second order: one from the binding
potential
\begin{eqnarray}
    \label{eq:potentialcorrection}
    v_{b}^{(2)}=& A& \int_{- \infty}^{+ \infty}dt  
    \int d^{3}{\bf q} e^{\frac{i\hbar t }{ 2m}(k^{2}-q^{2})}\\
     &\times& \frac{1 }{ N}\sum_{l}\left<b_{l}^{2}e^{i({\bf q}
     -{\bf k}){\bf R}_{l}(t)} 
     e^{i({\bf k}-{\bf q}){\bf R}_{l}(0)}\right> \nonumber
\end{eqnarray}
and the other from multiple scattering
\begin{eqnarray}
    \label{eq:mulscattcorrection}
    v_{m}^{(2)} = & A &\int_{- \infty}^{+ \infty} dt 
    \int d^{3}{\bf q}e^{\frac{i\hbar t }{ 2m}(k^{2}-q^{2})}\\
    & & \mbox{} \times    \frac{1}{ N} \sum_{l,l^{'}, l\ne l^{'}}    
    \left[
     \left<b_{l}b_{l^{'}}    e^{i({\bf q}-{\bf k}){\bf R}_{l}(t)}    
     e^{i({\bf k}-{\bf q}){\bf
     R}_{l^{'}}(0)}\right> \right.\nonumber\\
    & & \mbox{}\mbox{} -  \left< b_{l}    
    e^{ i ( {\bf q}-{\bf k} ){\bf R}_{l}(t)} \right> 
    \left.\left<b_{l^{'}}    e^{i({\bf k}-{\bf q}){\bf R}_{l^{'}}(0)}
    \right> \right]\nonumber
\end{eqnarray}

\noindent where $A=\frac{-i \rho }{ \left(2\pi\right)^{3} \hbar}  \left(\frac{2\pi \hbar^{2} }{m}\right)^{2}$. In these equations $\rho$ is the number density of scatterers, $k$
and $q$ are the wave vectors of the incident and in-medium
neutrons (very close to identical for cold neutrons), $b_{l}$ is
the (spin dependent) scattering length operator of atom $l$, and
the averaging $\left<\right>$ is the usual trace over spins and
internal states of the scattering system. These expressions make it clear that the second order
approximation to the optical potential is a function of the
dynamics and correlations of the scattering medium. The multiple
scattering term $v_{m}^{2}$ vanishes if the nuclei in the
medium possess uncorrelated nuclear spin directions
and uncorrelated relative motions, since multiple scattering
from uncorrelated nuclei cannot contribute to coherent scattering.

The parameters which control the size of these corrections are $kb$,  $kR$, and $b/d$ where $d$ is the separation between atoms in the medium~\cite{Sears1985, Sears1986}. For slow neutrons all of these parameters are typically of order $10^{-4}$. To our knowledge these corrections to the expression for the optical potential have not yet been measured experimentally. On neutron-nucleus resonances however in the eV energy range $kb \gg kR$ can approach values of order unity. In particular at a resonance energy $E_{r}$ one gets $kb={\Gamma_{n} / \Gamma_{tot}}$. This ratio approaches 1 if the inelastic contribution to the resonance is negligible compared to the elastic contribution. As will be seen below there are some neutron-nucleus resonances which come close to satisfying this condition. The inclusion of the effects beyond the kinematical theory of neutron optics in the calculation presented below is beyond the scope of this paper and does not modify any of the conclusions in this paper. 

If we write out the real and imaginary parts of the optical potential $U=V-iW$

\begin{equation}
    \label{eq:opticalpotential2}
    U=V-iW=(2\pi \hbar^2/m)\sum_{l} N_{l} (b_{l, r} -ib_{l, i}),
\end{equation}

\noindent and use this complex optical potential to calculate the reflection probability $|R^{2}|$ of a neutron incident on a uniform medium, one gets the result~\cite{Antonov69, Golub91}

\begin{equation}
    \label{eq:complexopticalpotential}
    |R^{2}|={{E_{\perp}-\sqrt{E_{\perp}(2\alpha-2(V-E_{\perp})}+\alpha}\over {E_{\perp}+\sqrt{E_{\perp}(2\alpha-2(V-E_{\perp})}+\alpha}}
\end{equation}

where

\begin{equation}
    \label{eq:opticalpotential3}
    \alpha=\sqrt{(V-E_{\perp})^{2}+W^{2}}
\end{equation}

\noindent and $E_{\perp}=\hbar^{2}k_{z}^{2}/2m$ is the fraction of the incident neutron kinetic energy associated with the momentum normal to the material boundary. 

In our case we want to consider the situation in which the neutron-nucleus scattering possesses both a potential scattering term as usual and also a resonance scattering term. The resonance both absorbs neutrons and also adds a large imaginary component to the total n-A neutron scattering amplitude.  In the presence of n-A resonances the expression for the resonant part $b_{res}$ of the total scattering amplitude $b=b_{pot}+b_{res}$ becomes~\cite{Mughabghab81}

\begin{equation}
    \label{eq:opticalpotential4}
    b_{res}=\sum_{j}{g_{\pm, j} \over 2k^{'}_{j}}{\Gamma_{n,j} \over [(E^{'}-E_{j})+i\Gamma_{j}/2]}
\end{equation}

\noindent where $\Gamma_{n,j}$ and $\Gamma_{j}$ are the neutron width and total width of the resonance at energy $E_{j}$ and $k^{'}=\mu k/m$ is the wave vector in the n-A center of mass system of reduced mass $\mu$,  $E^{'}$ is the associated energy in the COM frame, and $g_{+, j}=(I+1)/(2I+1)$ and $g_{-,j}=I/(2I+1)$ are the statistical weight factors for a resonance at energy $E_{j}$ in the total angular momentum channel $J=I \pm 1/2$. $b_{res}$ is purely imaginary on resonance, as can easily be seen explicitly by writing out the real and imaginary parts of $b_{res}$

\begin{equation}
    \label{eq:opticalpotential5}
    b_{res, real}=\sum_{j}{g_{\pm, j} \over 2k^{'}_{j}}{\Gamma_{n,j}(E^{'}-E_{j}) \over [(E^{'}-E_{j})^{2}+\Gamma_{j}^{2}/4]}
\end{equation}

\begin{equation}
    \label{eq:opticalpotential6}
b_{res, im}=-\sum_{j}{g_{\pm, j} \over 4k^{'}_{j}}{\Gamma_{n,j}\Gamma_{j} \over [(E^{'}-E_{j})^{2}+\Gamma_{j}^{2}/4]}
\end{equation}

\section{Analysis and Results}

\begin{figure*}
\centering
\includegraphics[scale=0.35]{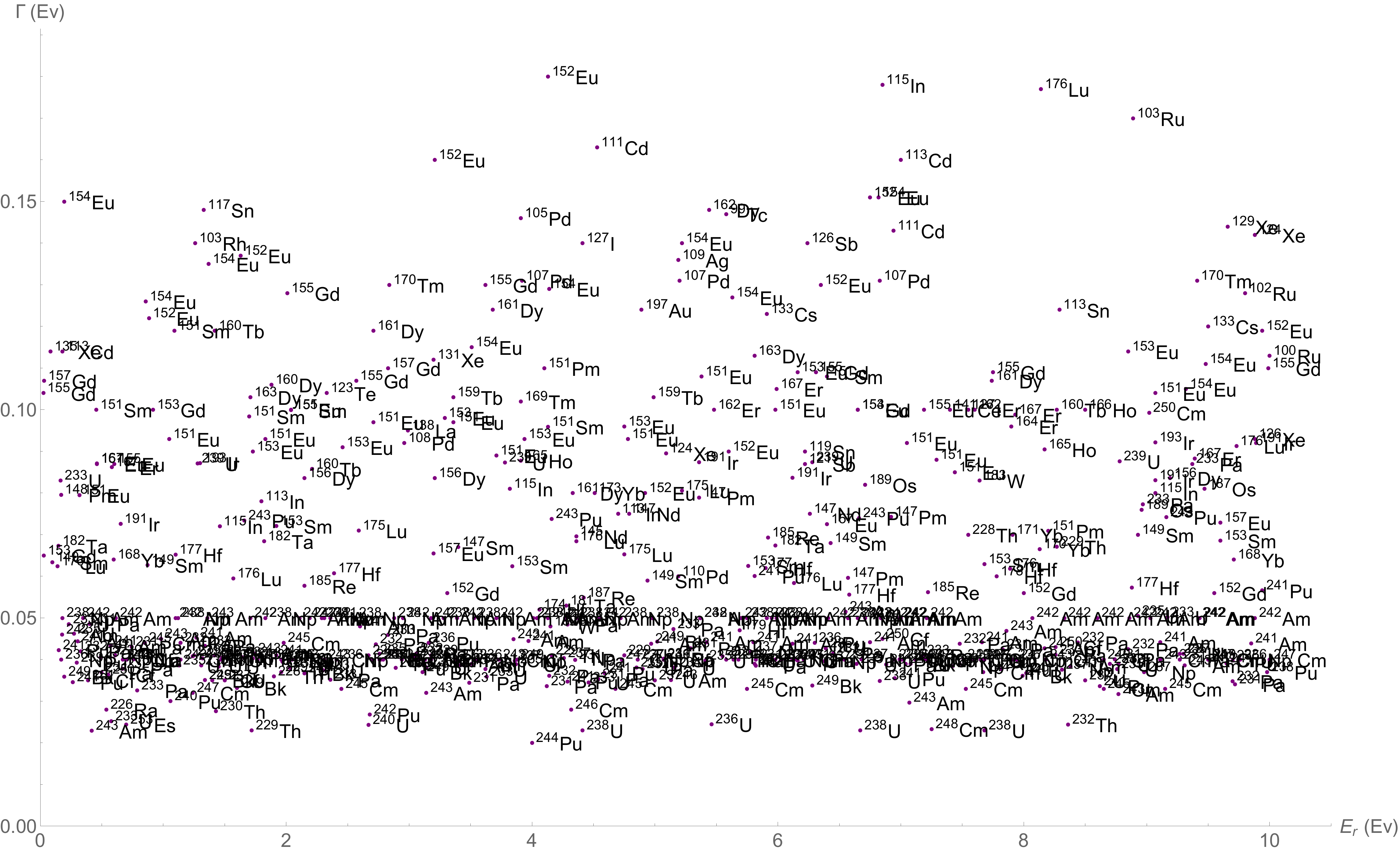}
\caption[]{Plot of total resonance widths of neutron resonances versus resonance energy with identification of the relevant isotopes.}
\label{fig:GammaEnergyIsotopes}
\end{figure*} 

\begin{figure*}
\centering
\includegraphics[scale=0.4]{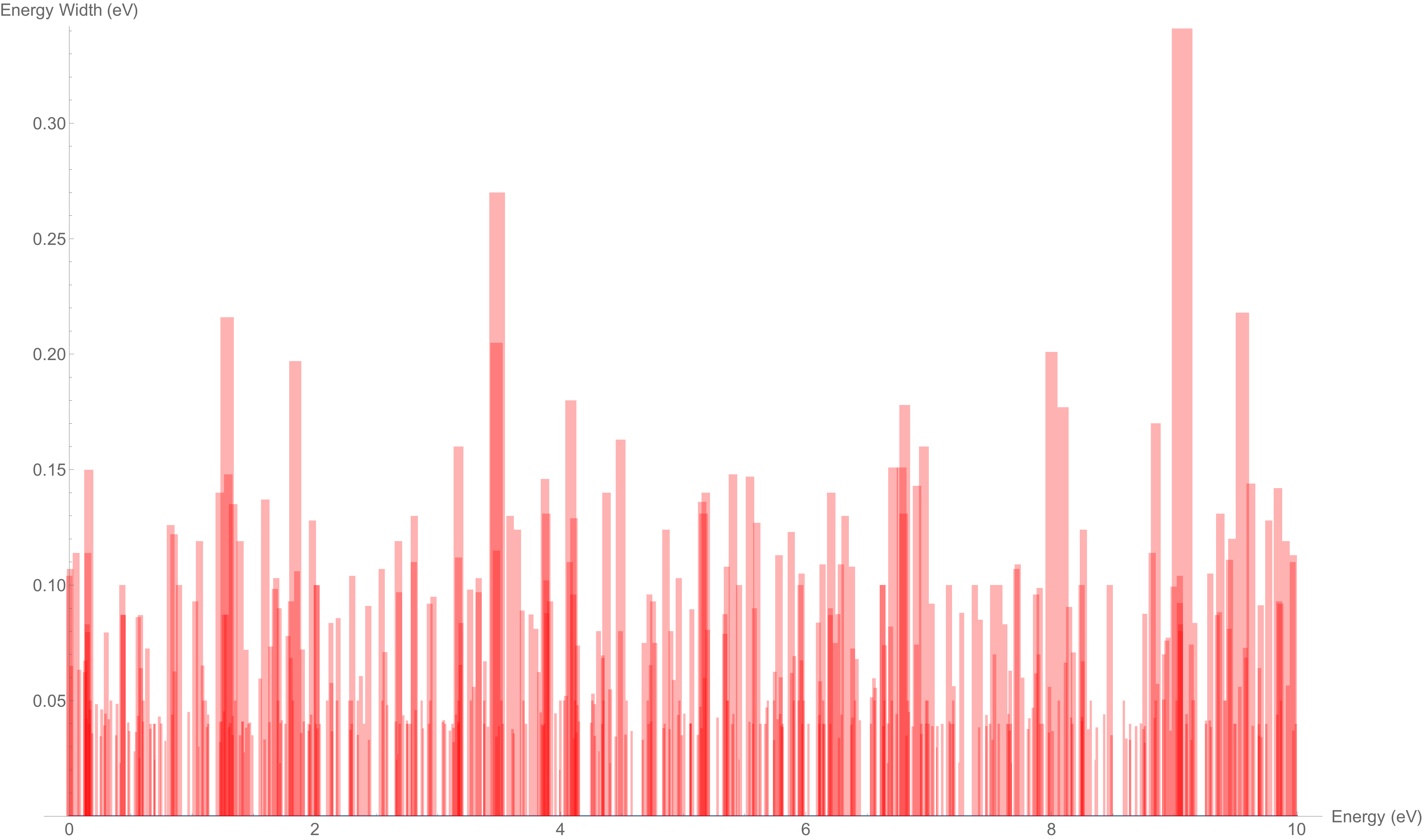}
\caption[]{Plot of total resonance widths of neutron resonances versus resonance energy for heavy nuclei in a representation that illustrates the \enquote{forest}. The horizontal dimension of the bars is $2\Gamma$.}
\label{fig:GammaEnergyLinePlot}
\end{figure*} 

Extensive data on n-A resonances exists. Table 1 in the Appendix lists the nuclei and the resonances which we considered as in principle available as possible eV neutron mirror materials. Figures~\ref{fig:GammaEnergyIsotopes} and~\ref{fig:GammaEnergyLinePlot} show the total resonance widths, resonance energies, and associated isotopes in our energy range of interest. The
density of n-A resonances in this regime along with their overlapping total widths encouraged us to pursue the calculation. We consider here heavy stable nuclei with n-A resonances with energies below 10 eV.  This data is taken from the National Nuclear Data Center (NNDC)~\cite{NNDC} and the Japanese Evaluated Nuclear Data Library (JENDL)~\cite{Shibata2011}. We use these measured resonance parameters and substitute them into Eq.~\ref{eq:complexopticalpotential}. We neglected the attenuation of the evanescent wave of the reflecting neutrons in the medium, the Doppler broadening of the resonances due to the motion of the atoms at finite temperature, the roughness of the mirror surface~\cite{Steyerl1972}, and multiple scattering. Some of these effects can be very important for a quantitative analysis of the reflectivity in specific cases. 

We show the results of our calculations in a series of plots in which we simply superpose all of the reflectivity results from mirrors made of the relevant pure substances. The reflectivity of any compound material would be the appropriate weighted sum according to equation 9. We chose to use as variables the total neutron energy $E$ and the incident neutron angle $\theta$. We chose a typical value of $\theta=1$ mrad for all of the plots: at constant $E$ the reflectivity is a sharply decreasing function of $\theta$ according to equation 15. 

In the regime of unit reflectivity shown in Figure 3, all of the heavy nuclei fall below natural nickel and therefore there is no special advantage that we can see to use heavy nuclei in a eV neutron guide. The imaginary parts of the optical potential change the shape of the reflectivity curve away from the usual shape given by the Fresnel formula.  

The $1/k$ factor in the resonance amplitude eventually reduces the reflectivity at high energy. Nevertheless as can be seen in Figure 4 certain nuclei possess a high enough reflectivity on resonance in the 1-10 eV energy range to be clearly visible for an incident angle of 1 mRad. This observation is in our judgement the most interesting result of our calculation. 

\begin{figure*}
\centering
\includegraphics[scale=0.15]{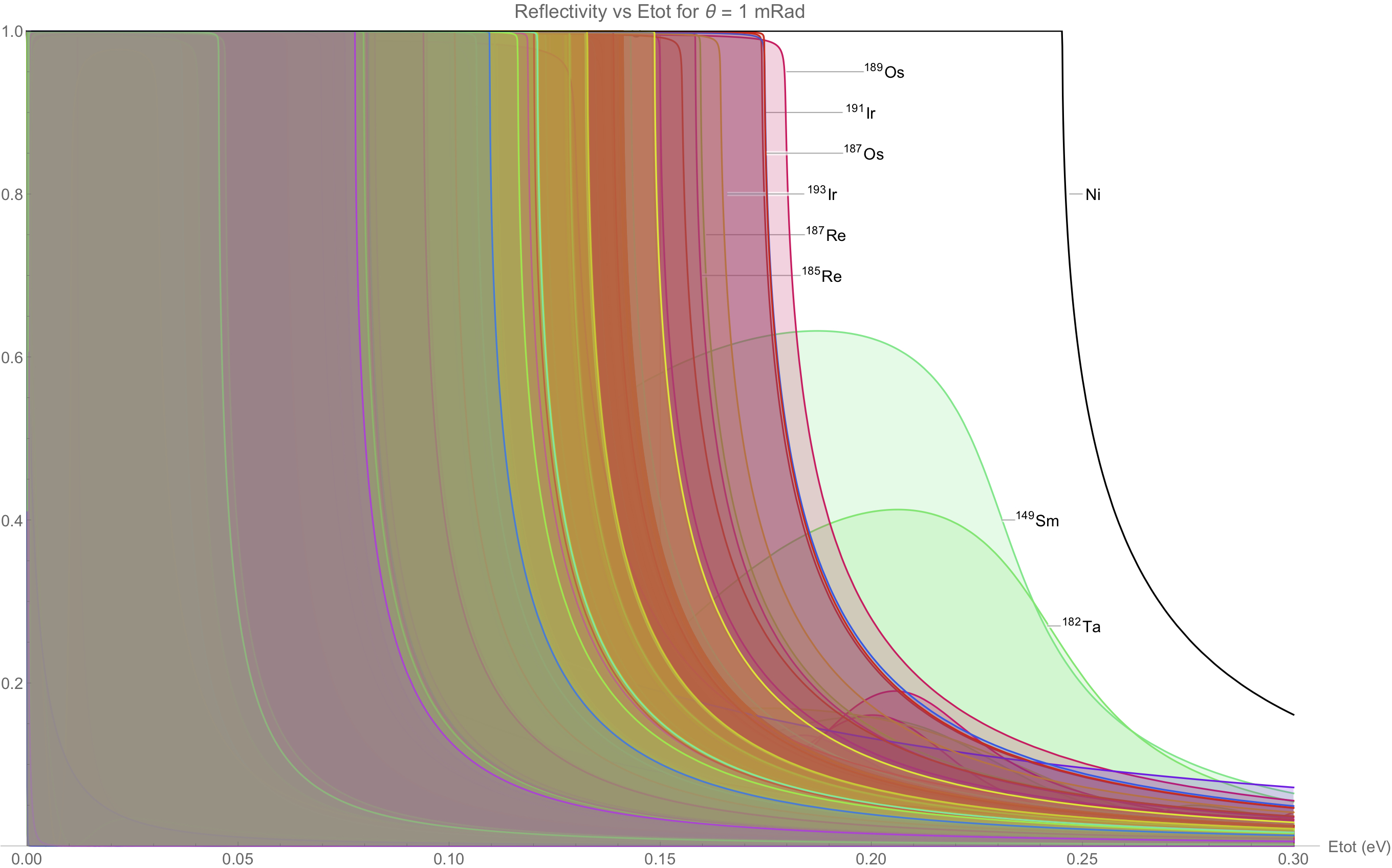}
\caption[]{Reflectivity for neutron energies between 0-0.3 eV for an incident angle of 1mRad. All of the heavy nuclei fall below natural nickel even after including the resonance contributions. The nuclei with reflectivity curves closest to natural nickel are labled explicitly.}
\label{fig:1mRad0-1eVreflectivity1}
\end{figure*} 

\begin{figure*}
\centering
\includegraphics[scale=0.5]{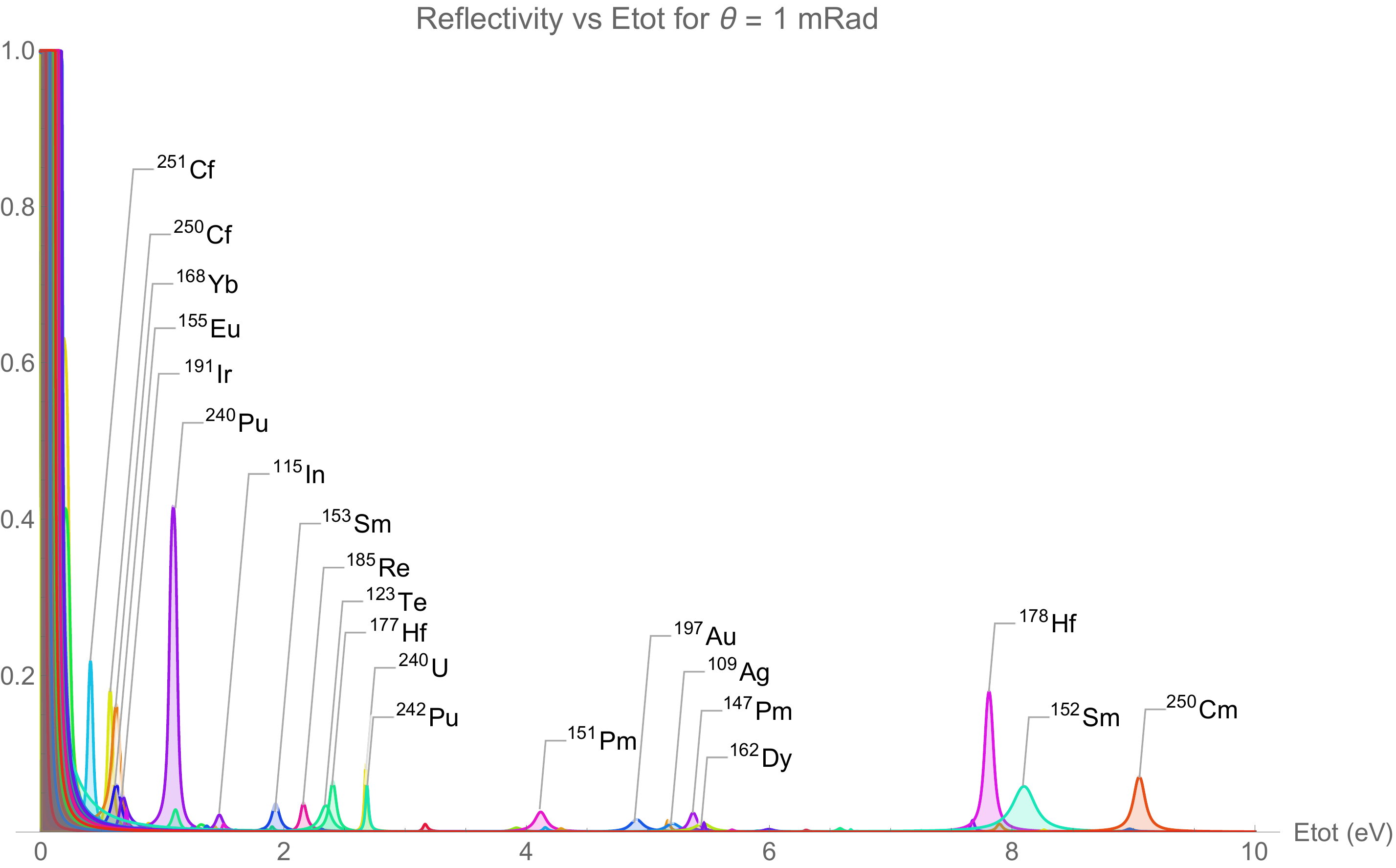}
\caption[]{Reflectivity envelope for the resonance guide for neutron energies between 0-10 eV for an incident angle of 1mRad.}
\label{fig:1mRad0-10eVreflectivity2}
\end{figure*} 

We included the effects of the real part of the scattering amplitudes as well in the formula for completeness. In principle one must take care not to double count the resonance scattering contribution, since the slow neutron scattering amplitudes $b_{measured}$ that are reported in the literature from measurements using slow neutrons are in fact a sum of the potential scattering contribution and also the tails of all of the other resonances in the limit $E \to 0$: 

\begin{equation}
    \label{eq:resonancetails}
b_{measured}=R+\sum_{j}{g_{\pm, j} \over 2k^{'}_{j}}{\Gamma_{n,j} \over [(E_{j})+i\Gamma_{j}/2]}\,\cdot
\end{equation} 

We estimated this effect but it is typically rather small and makes a negligible change in the reflectivity. We also estimated the effect of Doppler broadening of the resonances which would be present for a neutron mirror at room temperature~\cite{Lamb1939} and the lowered reflectivity of the mirror due to thermal diffuse scattering from phonons~\cite{Freund1983}. These effects are small enough that they do not modify any of our conclusions.

Finally we plot the reflectivity for a few special nuclei ($^{155}$Gd, $^{157}$Gd, $^{113}$Cd, and $^{10}$B) which have especially large absorption cross sections close to zero neutron energy from either a subthreshold resonance (in the case of $^{10}$B) or from an especially low energy resonance (as in $^{155}$Gd, $^{157}$Gd, and $^{113}$Cd). One can see that the reflectivity is reasonably large up to about 100 meV. This fact could conceivably be of interest for neutron beam transport in thermal and epithermal neutron beams in sections of the beamline close to the neutron source. Such a guide would both reflect some of the neutrons in the epithermal energy range and also efficiently absorb many of the neutrons incident on the guide above the critical angle.    

\begin{figure*}
\centering
\includegraphics[scale=0.4]{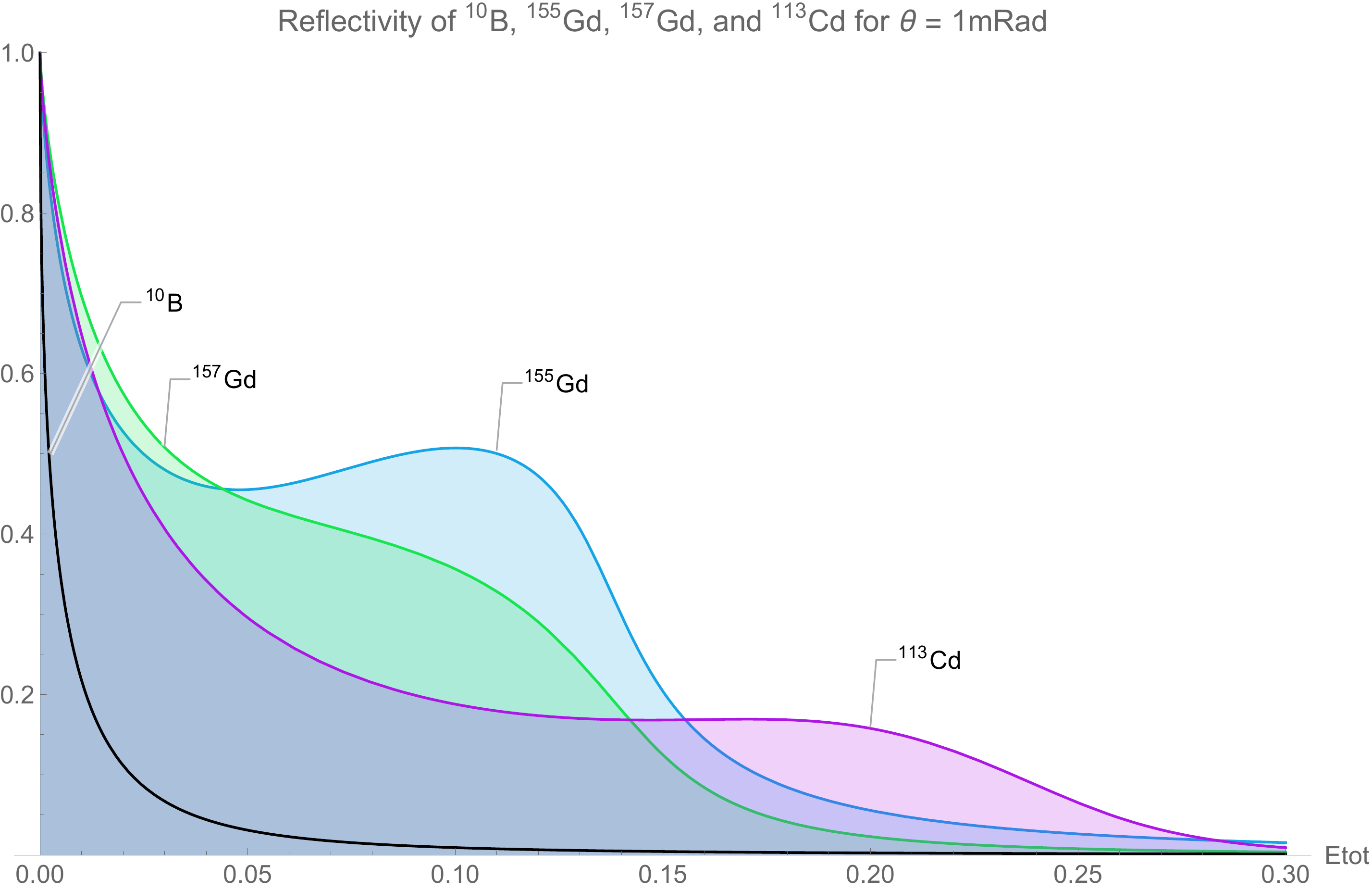}
\caption[]{Reflectivity of $^{155}$Gd, $^{157}$Gd, $^{113}$Cd, and $^{10}$B between 0-0.3 eV for an incident angle of 1mRad.}
\label{fig:1mRad0-10eVreflectivity3}
\end{figure*}

\section{Possible Applications}

Figure 4 shows that there are a reasonably large number of heavy nuclei which possess resonances whose properties can produce a noticeably large neutron reflectivity on resonance.  The relatively high neutron reflectivities for these cases opens the possibility for new applications in nondestructive subsurface analysis. One can exploit the relatively shallow penetration of the neutron wave amplitude into the medium upon mirror reflection to detect the presence of certain subsurface isotopes from the neutron energy dependence of the reflectivity from the flat surface of a material. As the reflected neutrons can be separated from the incident beam and this enhanced reflectivity is completely absent for all other nuclei, this method could possess a high signal to background ratio in a practical apparatus. One could also exploit the fact that the nuclear resonance will also emit a gamma cascade upon neutron capture to detect both this gamma cascade along with the delayed coincidence with the reflectivity peak at the resonance energy to further suppress environmental backgrounds. This latter mode of operation can easily be realized at a pulsed neutron source where the neutron energy is tightly correlated with neutron time of flight.        

Table 1 in the Appendix shows the product $kb={\Gamma_{n} / \Gamma_{tot}}$ for all of the resonances considered at the resonance peaks. Note that in some cases $kb$ is close to unity. In light of our earlier 
discussion on the regime of validity of the kinematical theory of neutron optics, it is clear that the neutron reflectivity from these resonances would be a sensitive observable with which to investigate the full dynamical theory of neutron optics, a subject which has yet to be explored experimentally. A more quantitative expression for the reflectivity from the resonances would need to take these higher order multiple scattering effects into account and would require knowledge of the structure and dynamics of the specific material under consideration. For the convenience of future researchers who might want to pursue such an investigation we have listed in the Appendix the relevant properties of the resonances.

\section{Conclusion}

We present a calculation of the reflectivity of neutron mirrors  in the eV energy range which takes into account the large imaginary parts of neutron-nucleus resonance scattering amplitudes based on the kinematic theory of neutron optics and the known properties of neutron-nucleus resonances. We find that the reflectivity is large enough to be visible for several nuclei on different resonances in the 0 to 10 eV neutron energy range. The relatively high neutron reflectivities for these cases opens the possibility for new applications in nondestructive subsurface analysis and for the investigation of multiple scattering effects in the theory of neutron optics. 

\section{Acknowledgements}

W. M. Snow acknowledges support from US National Science Foundation
grants PHY-1614545 and  PHY-1707988. W. M. Snow, K. A. Dickerson, J. S.  Devaney, and C. Haddock acknowledge support from the Indiana University Center for Spacetime Symmetries. W. M. Snow thanks the Neutron Physics Group in the Radiation Physics Division of the NIST Physical Measurement Laboratory in Gaithersburg, MD for the opportunity to present this work in preliminary form in a seminar. This work was completed during the ECT Workshop \enquote{Discrete Symmetries in Particle, Nuclear and Atomic Physics and implications for our Universe} held in Trento, Italy on Oct. 8-12, 2018. 

\section{Appendix}
\noindent
\begin{longtable*}[c]{ m{1.7cm}  m{2.3cm} m{2cm} m{2cm} m{2cm} m{2cm} m{2cm}}
\caption{A list of neutron-nucleus resonances considered in this paper with their respective resonance parameters and the product $kb$ of the resonance wave vector and the scattering amplitude on resonance.}
\label{my-label}
\\\hline\\
{\bf Isotope} &
{\bf Resonance Energy (eV)} & {\bf $\bf{\Gamma}$ (meV)} & {\bf $\bf{\Gamma_{n}}$ (meV)} &  $kb_{res}=\bf{\Gamma_{n}} / \bf{\Gamma}$\\
\endfirsthead
\\\hline\\
{\bf Isotope} & {\bf Resonance Energy (eV)} & {\bf $\bf{\Gamma}$ (meV)} & {\bf $\bf{\Gamma_{n}}$ (meV)} & $kb_{res}=\bf{\Gamma_{n}} / \bf{\Gamma}$ \\
\endhead
\\\hline\\
\endfoot
\\\hline\\
\endlastfoot\\
$^{87}$Sr  & 3.53    & 205   & 0.618      & 3.01E-03 \\
$^{99}$Tc  & 5.58    & 147   & 3.45       & 2.35E-02 \\
$^{100}$Ru & 10      & 113   & 0.138      & 1.22E-03 \\
$^{102}$Ru & 9.8     & 128   & 0.0093     & 7.27E-05 \\
$^{103}$Ru & 8.89    & 170   & 0.134      & 7.88E-04 \\
$^{103}$Rh & 1.26    & 140   & 0.52       & 3.71E-03 \\
$^{105}$Rh & 5       & 490   & 330        & 6.73E-01 \\
$^{105}$Pd & 3.91    & 146   & 0.000613   & 4.20E-06 \\
$^{107}$Pd & 3.92    & 131   & 0.00171    & 1.31E-05 \\
$^{107}$Pd & 5.2     & 131   & 0.0197     & 1.50E-04 \\
$^{107}$Pd & 6.83    & 131   & 0.0588     & 4.49E-04 \\
$^{108}$Pd & 2.96    & 92    & 0.01       & 1.09E-04 \\
$^{110}$Pd & 5.19    & 60    & 0.0024     & 4.00E-05 \\
$^{109}$Ag & 5.19    & 136   & 12.7       & 9.34E-02 \\
$^{111}$Cd & 4.53    & 163   & 0.0014     & 8.59E-06 \\
$^{111}$Cd & 6.94    & 143   & 0.00086    & 6.01E-06 \\
$^{113}$Cd & 0.179   & 114   & 0.64       & 5.61E-03 \\
$^{113}$Cd & 7       & 160   & 0.000413   & 2.58E-06 \\
$^{113}$In & 1.8     & 78    & 0.223      & 2.86E-03 \\
$^{113}$In & 4.7     & 75    & 0.106      & 1.41E-03 \\
$^{115}$In & 1.46    & 72    & 2.98       & 4.14E-02 \\
$^{115}$In & 3.82    & 81    & 0.38       & 4.69E-03 \\
$^{115}$In & 6.85    & 178   & 0.000418   & 2.35E-06 \\
$^{115}$In & 9.07    & 80    & 1.5        & 1.88E-02 \\
$^{113}$Sn & 8.29    & 124   & 4.47       & 3.60E-02 \\
$^{117}$Sn & 1.33    & 148   & 0.00011    & 7.43E-07 \\
$^{119}$Sn & 6.22    & 90    & 0.00148    & 1.64E-05 \\
$^{121}$Sb & 6.22    & 87    & 1.88       & 2.16E-02 \\
$^{126}$Sb & 6.24    & 140   & 2.23       & 1.59E-02 \\
$^{123}$Te & 2.33    & 104   & 10.4       & 1.00E-01 \\
$^{127}$I  & 4.41    & 140   & 0.0001     & 7.14E-07 \\
$^{127}$I  & 7.55    & 100   & 0.000483   & 4.83E-06 \\
$^{124}$Xe & 5.09    & 89.5  & 12.9       & 1.44E-01 \\
$^{124}$Xe & 9.88    & 142   & 42.3       & 2.98E-01 \\
$^{126}$Xe & 9.88    & 93    & 1.35       & 1.45E-02 \\
$^{129}$Xe & 9.66    & 144   & 7.76       & 5.39E-02 \\
$^{131}$Xe & 3.2     & 112   & 0.000256   & 2.29E-06 \\
$^{135}$Xe & 0.0842  & 114   & 20.4       & 1.79E-01 \\
$^{133}$Cs & 5.91    & 123   & 7.38       & 6.00E-02 \\
$^{133}$Cs & 9.5     & 120   & 0.00217    & 1.81E-05 \\
$^{138}$La & 2.99    & 95    & 1.19       & 1.25E-02 \\
$^{139}$La & 0.758   & 40.1  & 0.0000996  & 2.48E-06 \\
$^{141}$Ce & 7.4     & 100   & 1.01       & 1.01E-02 \\
$^{145}$Nd & 4.36    & 69.6  & 1.18       & 1.70E-02 \\
$^{147}$Nd & 4.79    & 75    & 0.309      & 4.12E-03 \\
$^{147}$Nd & 6.26    & 75    & 0.651      & 8.68E-03 \\
$^{147}$Pm & 5.36    & 78.9  & 40.1       & 5.08E-01 \\
$^{147}$Pm & 6.57    & 59.6  & 1.4        & 2.35E-02 \\
$^{147}$Pm & 6.92    & 74.3  & 4.75       & 6.39E-02 \\
$^{148}$Pm & 0.169   & 79.6  & 0.545      & 6.85E-03 \\
$^{151}$Pm & 4.1     & 110   & 41.4       & 3.76E-01 \\
$^{151}$Pm & 8.2     & 70.9  & 1.87       & 2.64E-02 \\
$^{147}$Sm & 3.4     & 67    & 1.35       & 2.01E-02 \\
$^{149}$Sm & 0.0973  & 63.4  & 0.525      & 8.28E-03 \\
$^{149}$Sm & 0.872   & 62.7  & 0.75       & 1.20E-02 \\
$^{149}$Sm & 4.94    & 59    & 2.14       & 3.63E-02 \\
$^{149}$Sm & 6.43    & 68    & 1.2        & 1.76E-02 \\
$^{149}$Sm & 8.93    & 70    & 11.8       & 1.69E-01 \\
$^{151}$Sm & 0.456   & 100   & 0.0223     & 2.23E-04 \\
$^{151}$Sm & 1.09    & 119   & 0.948      & 7.97E-03 \\
$^{151}$Sm & 1.7     & 98.4  & 0.444      & 4.51E-03 \\
$^{151}$Sm & 2.04    & 99.9  & 0.744      & 7.45E-03 \\
$^{151}$Sm & 4.13    & 95.9  & 1.26       & 1.31E-02 \\
$^{151}$Sm & 6.4     & 108   & 6.71       & 6.21E-02 \\
$^{152}$Sm & 8.05    & 201   & 135        & 6.72E-01 \\
$^{153}$Sm & 1.92    & 72.1  & 10.1       & 1.40E-01 \\
$^{153}$Sm & 3.84    & 62.4  & 0.42       & 6.73E-03 \\
$^{153}$Sm & 5.76    & 62.5  & 0.523      & 8.37E-03 \\
$^{153}$Sm & 7.68    & 62.9  & 0.917      & 1.46E-02 \\
$^{153}$Sm & 9.6     & 68.6  & 6.61       & 9.64E-02 \\
$^{151}$Eu & 0.32    & 79.5  & 0.071      & 8.93E-04 \\
$^{151}$Eu & 0.46    & 87    & 0.665      & 7.64E-03 \\
$^{151}$Eu & 1.05    & 93    & 0.19       & 2.04E-03 \\
$^{151}$Eu & 1.83    & 93    & 0.0324     & 3.48E-04 \\
$^{151}$Eu & 2.71    & 97    & 0.223      & 2.30E-03 \\
$^{151}$Eu & 3.36    & 97    & 1.1        & 1.13E-02 \\
$^{151}$Eu & 3.71    & 89    & 0.694      & 7.80E-03 \\
$^{151}$Eu & 4.78    & 93    & 0.146      & 1.57E-03 \\
$^{151}$Eu & 5.38    & 108   & 0.228      & 2.11E-03 \\
$^{151}$Eu & 5.98    & 100   & 0.408      & 4.08E-03 \\
$^{151}$Eu & 7.05    & 92    & 0.0429     & 4.66E-04 \\
$^{151}$Eu & 7.29    & 88    & 1.89       & 2.15E-02 \\
$^{151}$Eu & 7.44    & 85    & 2.03       & 2.39E-02 \\
$^{151}$Eu & 9.07    & 104   & 1.3        & 1.25E-02 \\
$^{152}$Eu & 0.884   & 122   & 0.187      & 1.53E-03 \\
$^{152}$Eu & 1.34    & 216   & 0.146      & 6.76E-04 \\
$^{152}$Eu & 1.63    & 137   & 0.104      & 7.59E-04 \\
$^{152}$Eu & 1.89    & 197   & 0.48       & 2.44E-03 \\
$^{152}$Eu & 3.21    & 160   & 0.0035     & 2.19E-05 \\
$^{152}$Eu & 3.55    & 270   & 0.791      & 2.93E-03 \\
$^{152}$Eu & 4.13    & 180   & 0.391      & 2.17E-03 \\
$^{152}$Eu & 4.92    & 80    & 0.285      & 3.56E-03 \\
$^{152}$Eu & 5.6     & 90    & 0.199      & 2.21E-03 \\
$^{152}$Eu & 6.35    & 130   & 0.551      & 4.24E-03 \\
$^{152}$Eu & 6.75    & 151   & 0.543      & 3.60E-03 \\
$^{152}$Eu & 9.61    & 218   & 2.57       & 1.18E-02 \\
$^{152}$Eu & 9.94    & 119   & 0.665      & 5.59E-03 \\
$^{153}$Eu & 1.73    & 90    & 0.0489     & 5.43E-04 \\
$^{153}$Eu & 2.46    & 91    & 1.03       & 1.13E-02 \\
$^{153}$Eu & 3.29    & 98    & 1.09       & 1.11E-02 \\
$^{153}$Eu & 3.94    & 93    & 0.943      & 1.01E-02 \\
$^{153}$Eu & 4.75    & 96    & 0.0429     & 4.47E-04 \\
$^{153}$Eu & 6.16    & 109   & 0.54       & 4.95E-03 \\
$^{153}$Eu & 8.85    & 114   & 3          & 2.63E-02 \\
$^{154}$Eu & 0.195   & 150   & 0.0717     & 4.78E-04 \\
$^{154}$Eu & 0.857   & 126   & 0.056      & 4.44E-04 \\
$^{154}$Eu & 1.37    & 135   & 0.461      & 3.41E-03 \\
$^{154}$Eu & 3.51    & 115   & 0.222      & 1.93E-03 \\
$^{154}$Eu & 4.14    & 129   & 0.642      & 4.98E-03 \\
$^{154}$Eu & 5.22    & 140   & 0.359      & 2.56E-03 \\
$^{154}$Eu & 5.63    & 127   & 0.227      & 1.79E-03 \\
$^{154}$Eu & 6.65    & 100   & 0.0303     & 3.03E-04 \\
$^{154}$Eu & 6.82    & 151   & 1.08       & 7.15E-03 \\
$^{154}$Eu & 9.32    & 105   & 3.16       & 3.01E-02 \\
$^{154}$Eu & 9.48    & 111   & 1.53       & 1.38E-02 \\
$^{155}$Eu & 0.602   & 87    & 4.08       & 4.69E-02 \\
$^{155}$Eu & 2.04    & 100   & 0.0394     & 3.94E-04 \\
$^{155}$Eu & 7.19    & 100   & 0.18       & 1.80E-03 \\
$^{157}$Eu & 3.2     & 65.5  & 0.504      & 7.69E-03 \\
$^{157}$Eu & 6.4     & 72.5  & 7.54       & 1.04E-01 \\
$^{157}$Eu & 9.6     & 72.9  & 7.92       & 1.09E-01 \\
$^{152}$Gd & 3.31    & 56    & 0.018      & 3.21E-04 \\
$^{152}$Gd & 8       & 56    & 5.1        & 9.11E-02 \\
$^{152}$Gd & 9.55    & 56    & 0.093      & 1.66E-03 \\
$^{153}$Gd & 0.0297  & 65    & 0.041      & 6.31E-04 \\
$^{153}$Gd & 0.917   & 100   & 0.0208     & 2.08E-04 \\
$^{153}$Gd & 6.65    & 100   & 1.04       & 1.04E-02 \\
$^{155}$Gd & 0.0252  & 104   & 0.097      & 9.33E-04 \\
$^{155}$Gd & 2.01    & 128   & 0.4        & 3.13E-03 \\
$^{155}$Gd & 2.57    & 107   & 1.71       & 1.60E-02 \\
$^{155}$Gd & 3.62    & 130   & 0.05       & 3.85E-04 \\
$^{155}$Gd & 6.31    & 109   & 2.2        & 2.02E-02 \\
$^{155}$Gd & 7.75    & 109   & 1.16       & 1.06E-02 \\
$^{155}$Gd & 9.99    & 110   & 0.2        & 1.82E-03 \\
$^{157}$Gd & 0.032   & 107   & 0.428      & 4.00E-03 \\
$^{157}$Gd & 2.83    & 110   & 0.377      & 3.43E-03 \\
$^{159}$Tb & 3.36    & 103   & 0.337      & 3.27E-03 \\
$^{159}$Tb & 4.99    & 103   & 0.0834     & 8.10E-04 \\
$^{160}$Tb & 1.42    & 119   & 3          & 2.52E-02 \\
$^{160}$Tb & 2.21    & 85.8  & 0.583      & 6.79E-03 \\
$^{160}$Tb & 8.27    & 100   & 8.72       & 8.72E-02 \\
$^{156}$Dy & 2.15    & 83.6  & 0.2        & 2.39E-03 \\
$^{156}$Dy & 3.21    & 83.6  & 0.8        & 9.57E-03 \\
$^{156}$Dy & 9.19    & 83.6  & 0.6        & 7.18E-03 \\
$^{160}$Dy & 1.88    & 106   & 0.2        & 1.89E-03 \\
$^{161}$Dy & 2.71    & 119   & 0.561      & 4.71E-03 \\
$^{161}$Dy & 3.68    & 124   & 2.14       & 1.73E-02 \\
$^{161}$Dy & 4.33    & 80    & 1.38       & 1.73E-02 \\
$^{161}$Dy & 7.74    & 107   & 0.514      & 4.80E-03 \\
$^{162}$Dy & 5.44    & 148   & 21         & 1.42E-01 \\
$^{163}$Dy & 1.71    & 103   & 2.04       & 1.98E-02 \\
$^{163}$Dy & 5.81    & 113   & 0.0231     & 2.04E-04 \\
$^{165}$Ho & 3.91    & 87.8  & 2.13       & 2.43E-02 \\
$^{165}$Ho & 8.17    & 90.5  & 0.187      & 2.07E-03 \\
$^{166}$Ho & 8.5     & 100   & 11         & 1.10E-01 \\
$^{162}$Er & 5.48    & 100   & 0.33       & 3.30E-03 \\
$^{162}$Er & 7.6     & 100   & 0.66       & 6.60E-03 \\
$^{164}$Er & 7.9     & 96    & 0.71       & 7.40E-03 \\
$^{167}$Er & 0.46    & 87.1  & 0.269      & 3.09E-03 \\
$^{167}$Er & 0.583   & 86.2  & 0.247      & 2.87E-03 \\
$^{167}$Er & 5.99    & 105   & 20.7       & 1.97E-01 \\
$^{167}$Er & 7.93    & 98.8  & 0.16       & 1.62E-03 \\
$^{167}$Er & 9.39    & 88.3  & 9.2        & 1.04E-01 \\
$^{168}$Tm & 0.00001 & 40000 & 1190       & 2.98E-02 \\
$^{169}$Tm & 3.91    & 102   & 7.47       & 7.32E-02 \\
$^{170}$Tm & 2.84    & 130   & 0.375      & 2.88E-03 \\
$^{170}$Tm & 9.41    & 131   & 1.25       & 9.54E-03 \\
$^{168}$Yb & 0.597   & 64    & 2.2        & 3.44E-02 \\
$^{168}$Yb & 9.71    & 64    & 0.08       & 1.25E-03 \\
$^{170}$Yb & 8.13    & 66.5  & 1.64       & 2.47E-02 \\
$^{171}$Yb & 7.91    & 70    & 2.2        & 3.14E-02 \\
$^{173}$Yb & 4.51    & 80    & 0.18       & 2.25E-03 \\
$^{175}$Lu & 2.59    & 71    & 0.188      & 2.65E-03 \\
$^{175}$Lu & 4.75    & 65.3  & 0.267      & 4.09E-03 \\
$^{175}$Lu & 5.22    & 80.6  & 1.6        & 1.99E-02 \\
$^{176}$Lu & 0.141   & 62.4  & 0.0927     & 1.49E-03 \\
$^{176}$Lu & 1.57    & 59.5  & 0.485      & 8.15E-03 \\
$^{176}$Lu & 4.36    & 68.4  & 0.403      & 5.89E-03 \\
$^{176}$Lu & 6.13    & 58.4  & 1.37       & 2.35E-02 \\
$^{176}$Lu & 8.14    & 177   & 0.0308     & 1.74E-04 \\
$^{176}$Lu & 9.73    & 91.3  & 1.28       & 1.40E-02 \\
$^{174}$Hf & 4.06    & 52    & 0.015      & 2.88E-04 \\
$^{176}$Hf & 7.89    & 61.8  & 10.1       & 1.63E-01 \\
$^{177}$Hf & 1.1     & 65.2  & 2.22       & 3.40E-02 \\
$^{177}$Hf & 2.39    & 60.7  & 8.04       & 1.32E-01 \\
$^{177}$Hf & 5.9     & 62    & 5.32       & 8.58E-02 \\
$^{177}$Hf & 6.58    & 55.6  & 8.21       & 1.48E-01 \\
$^{177}$Hf & 8.88    & 57.3  & 5.89       & 1.03E-01 \\
$^{178}$Hf & 7.78    & 60    & 50         & 8.33E-01 \\
$^{179}$Hf & 5.69    & 47    & 4.27       & 9.09E-02 \\
$^{181}$Ta & 4.28    & 53    & 3.2        & 6.04E-02 \\
$^{182}$Ta & 0.147   & 67.3  & 0.315      & 4.68E-03 \\
$^{182}$Ta & 1.82    & 68.4  & 1.35       & 1.97E-02 \\
$^{182}$Ta & 5.98    & 67.4  & 0.406      & 6.02E-03 \\
$^{182}$W  & 4.15    & 48    & 1.55       & 3.23E-02 \\
$^{183}$W  & 7.64    & 83    & 1.66       & 2.00E-02 \\
$^{185}$Re & 2.15    & 57.7  & 2.83       & 4.90E-02 \\
$^{185}$Re & 5.92    & 69.3  & 0.264      & 3.81E-03 \\
$^{185}$Re & 7.22    & 56.2  & 1.19       & 2.12E-02 \\
$^{187}$Re & 4.42    & 54.9  & 0.318      & 5.79E-03 \\
$^{187}$Os & 9.47    & 81    & 1.74       & 2.15E-02 \\
$^{189}$Os & 6.71    & 82    & 4.53       & 5.52E-02 \\
$^{189}$Os & 8.96    & 76    & 9.04       & 1.19E-01 \\
$^{191}$Ir & 0.653   & 72.6  & 0.44       & 6.06E-03 \\
$^{191}$Ir & 5.36    & 87.4  & 5.44       & 6.22E-02 \\
$^{191}$Ir & 6.12    & 83.7  & 0.707      & 8.45E-03 \\
$^{191}$Ir & 9.07    & 83.1  & 3.12       & 3.75E-02 \\
$^{191}$Ir & 9.89    & 92    & 1          & 1.09E-02 \\
$^{193}$Ir & 1.3     & 87.2  & 0.73       & 8.37E-03 \\
$^{193}$Ir & 9.07    & 92.2  & 2.24       & 2.43E-02 \\
$^{197}$Au & 4.89    & 124   & 15.2       & 1.23E-01 \\
$^{226}$Ra & 0.537   & 28    & 0.021      & 7.50E-04 \\
$^{228}$Th & 1.9     & 36    & 0.785      & 2.18E-02 \\
$^{228}$Th & 7.55    & 70    & 1.21       & 1.73E-02 \\
$^{229}$Th & 0.609   & 41    & 0.158      & 3.85E-03 \\
$^{229}$Th & 1.25    & 41    & 0.16       & 3.90E-03 \\
$^{229}$Th & 1.42    & 41    & 0.01       & 2.44E-04 \\
$^{229}$Th & 1.72    & 23    & 0.0643     & 2.80E-03 \\
$^{229}$Th & 1.96    & 37    & 0.223      & 6.03E-03 \\
$^{229}$Th & 2.67    & 41    & 0.0137     & 3.34E-04 \\
$^{229}$Th & 3.18    & 41    & 0.12       & 2.93E-03 \\
$^{229}$Th & 4.16    & 41    & 1.17       & 2.85E-02 \\
$^{229}$Th & 4.75    & 41    & 0.0317     & 7.73E-04 \\
$^{229}$Th & 5.58    & 40    & 0.497      & 1.24E-02 \\
$^{229}$Th & 6.95    & 40    & 0.912      & 2.28E-02 \\
$^{229}$Th & 8.27    & 67.1  & 0.137      & 2.04E-03 \\
$^{229}$Th & 9.15    & 341   & 0.432      & 1.27E-03 \\
$^{230}$Th & 1.43    & 27.6  & 0.39       & 1.41E-02 \\
$^{232}$Th & 8.36    & 24.4  & 0.000249   & 1.02E-05 \\
$^{231}$Pa & 0.4     & 48.6  & 0.0592     & 1.22E-03 \\
$^{231}$Pa & 0.497   & 36.9  & 0.0107     & 2.90E-04 \\
$^{231}$Pa & 0.744   & 43.1  & 0.0125     & 2.90E-04 \\
$^{231}$Pa & 1.24    & 40.8  & 0.0242     & 5.93E-04 \\
$^{231}$Pa & 1.96    & 40    & 0.0144     & 3.60E-04 \\
$^{231}$Pa & 2.79    & 40    & 0.013      & 3.25E-04 \\
$^{231}$Pa & 3.49    & 34.4  & 0.0386     & 1.12E-03 \\
$^{231}$Pa & 4.12    & 33.6  & 0.0709     & 2.11E-03 \\
$^{231}$Pa & 4.35    & 39.9  & 0.0557     & 1.40E-03 \\
$^{231}$Pa & 4.54    & 35.3  & 0.0202     & 5.72E-04 \\
$^{231}$Pa & 5.07    & 40.2  & 0.418      & 1.04E-02 \\
$^{231}$Pa & 5.29    & 42.7  & 0.0915     & 2.14E-03 \\
$^{231}$Pa & 5.64    & 40    & 0.0506     & 1.27E-03 \\
$^{231}$Pa & 5.82    & 40    & 0.0515     & 1.29E-03 \\
$^{231}$Pa & 6.21    & 40    & 0.052      & 1.30E-03 \\
$^{231}$Pa & 6.54    & 40    & 0.0707     & 1.77E-03 \\
$^{231}$Pa & 6.88    & 38.8  & 0.21       & 5.41E-03 \\
$^{231}$Pa & 7.58    & 40    & 0.147      & 3.68E-03 \\
$^{231}$Pa & 7.83    & 42.4  & 0.261      & 6.16E-03 \\
$^{231}$Pa & 8.74    & 40.1  & 1.08       & 2.69E-02 \\
$^{231}$Pa & 9.27    & 40    & 0.056      & 1.40E-03 \\
$^{231}$Pa & 9.72    & 34.1  & 0.349      & 1.02E-02 \\
$^{232}$Pa & 0.33    & 42    & 0.0676     & 1.61E-03 \\
$^{232}$Pa & 0.67    & 37.8  & 0.0371     & 9.81E-04 \\
$^{232}$Pa & 1.14    & 44    & 0.0348     & 7.91E-04 \\
$^{232}$Pa & 1.42    & 41    & 0.0223     & 5.44E-04 \\
$^{232}$Pa & 2.73    & 43.7  & 0.457      & 1.05E-02 \\
$^{232}$Pa & 3.06    & 41.5  & 0.344      & 8.29E-03 \\
$^{232}$Pa & 4.14    & 36.1  & 0.484      & 1.34E-02 \\
$^{232}$Pa & 4.9     & 37.7  & 0.179      & 4.75E-03 \\
$^{232}$Pa & 5.55    & 40.8  & 0.0462     & 1.13E-03 \\
$^{232}$Pa & 5.82    & 38.5  & 0.138      & 3.58E-03 \\
$^{232}$Pa & 6.45    & 41.5  & 0.505      & 1.22E-02 \\
$^{232}$Pa & 7.48    & 43.8  & 0.144      & 3.29E-03 \\
$^{232}$Pa & 8.44    & 44.1  & 0.338      & 7.66E-03 \\
$^{232}$Pa & 8.85    & 42.5  & 0.507      & 1.19E-02 \\
$^{232}$Pa & 9.7     & 34.8  & 0.433      & 1.24E-02 \\
$^{233}$Pa & 0.789   & 32.6  & 0.00144    & 4.42E-05 \\
$^{233}$Pa & 1.34    & 43.3  & 0.108      & 2.49E-03 \\
$^{233}$Pa & 1.64    & 40.6  & 0.309      & 7.61E-03 \\
$^{233}$Pa & 2.36    & 35.2  & 0.00744    & 2.11E-04 \\
$^{233}$Pa & 2.83    & 45.9  & 0.156      & 3.40E-03 \\
$^{233}$Pa & 3.39    & 40    & 0.526      & 1.32E-02 \\
$^{233}$Pa & 4.29    & 48.4  & 0.0876     & 1.81E-03 \\
$^{233}$Pa & 5.15    & 47.3  & 0.399      & 8.44E-03 \\
$^{233}$Pa & 7.18    & 41    & 0.139      & 3.39E-03 \\
$^{233}$Pa & 8.26    & 41.2  & 0.0401     & 9.73E-04 \\
$^{233}$Pa & 8.97    & 77.3  & 0.158      & 2.04E-03 \\
$^{233}$Pa & 9.37    & 87    & 1.05       & 1.21E-02 \\
$^{232}$U  & 5.98    & 40    & 1.5        & 3.75E-02 \\
$^{233}$U  & 0.166   & 83    & 0.0001     & 1.20E-06 \\
$^{233}$U  & 0.232   & 48.5  & 0.0000318  & 6.56E-07 \\
$^{233}$U  & 0.577   & 25.2  & 0.000629   & 2.50E-05 \\
$^{233}$U  & 1.45    & 38.3  & 0.201      & 5.25E-03 \\
$^{233}$U  & 1.77    & 40    & 0.258      & 6.45E-03 \\
$^{233}$U  & 2.3     & 37.4  & 0.152      & 4.06E-03 \\
$^{233}$U  & 3.51    & 39    & 0.181      & 4.64E-03 \\
$^{233}$U  & 3.63    & 35.9  & 0.0545     & 1.52E-03 \\
$^{233}$U  & 4.46    & 34.5  & 0.379      & 1.10E-02 \\
$^{233}$U  & 5.81    & 39    & 0.1        & 2.56E-03 \\
$^{233}$U  & 6.64    & 39    & 0.343      & 8.79E-03 \\
$^{233}$U  & 6.83    & 34.9  & 0.689      & 1.97E-02 \\
$^{233}$U  & 7.48    & 39    & 0.0334     & 8.56E-04 \\
$^{233}$U  & 8.7     & 39    & 0.019      & 4.87E-04 \\
$^{233}$U  & 8.77    & 39    & 0.329      & 8.44E-03 \\
$^{233}$U  & 9.17    & 50.2  & 0.0677     & 1.35E-03 \\
$^{234}$U  & 5.16    & 38.1  & 3.64       & 9.55E-02 \\
$^{235}$U  & 0.274   & 46.2  & 0.00425    & 9.20E-05 \\
$^{235}$U  & 1.13    & 38.6  & 0.0145     & 3.76E-04 \\
$^{235}$U  & 1.31    & 38.6  & 0.000195   & 5.05E-06 \\
$^{235}$U  & 2.03    & 38    & 0.009      & 2.37E-04 \\
$^{235}$U  & 2.76    & 41.4  & 0.000738   & 1.78E-05 \\
$^{235}$U  & 3.14    & 38.2  & 0.0238     & 6.23E-04 \\
$^{235}$U  & 3.62    & 37.7  & 0.0421     & 1.12E-03 \\
$^{235}$U  & 3.87    & 38.9  & 0.000445   & 1.14E-05 \\
$^{235}$U  & 4.85    & 38.1  & 0.0525     & 1.38E-03 \\
$^{235}$U  & 5.41    & 39.8  & 0.022      & 5.53E-04 \\
$^{235}$U  & 6.16    & 39.8  & 0.0533     & 1.34E-03 \\
$^{235}$U  & 6.39    & 42.7  & 0.242      & 5.67E-03 \\
$^{235}$U  & 6.99    & 40.1  & 0.00156    & 3.89E-05 \\
$^{235}$U  & 7.08    & 39    & 0.112      & 2.87E-03 \\
$^{235}$U  & 7.66    & 39.8  & 0.00202    & 5.08E-05 \\
$^{235}$U  & 8.76    & 37.7  & 0.966      & 2.56E-02 \\
$^{235}$U  & 8.93    & 50.5  & 0.0829     & 1.64E-03 \\
$^{235}$U  & 9.27    & 41.4  & 0.12       & 2.90E-03 \\
$^{235}$U  & 9.7     & 39.8  & 0.0356     & 8.94E-04 \\
$^{236}$U  & 5.46    & 24.5  & 2.3        & 9.39E-02 \\
$^{237}$U  & 1.5     & 35    & 0.533      & 1.52E-02 \\
$^{237}$U  & 5       & 35    & 1.04       & 2.97E-02 \\
$^{237}$U  & 8.5     & 35    & 1.36       & 3.89E-02 \\
$^{238}$U  & 4.41    & 23    & 0.0000555  & 2.41E-06 \\
$^{238}$U  & 6.67    & 23    & 1.48       & 6.43E-02 \\
$^{238}$U  & 7.68    & 23    & 0.00000942 & 4.10E-07 \\
$^{239}$U  & 1.28    & 87.1  & 0.282      & 3.24E-03 \\
$^{239}$U  & 3.78    & 87.3  & 0.486      & 5.57E-03 \\
$^{239}$U  & 6.28    & 87.4  & 0.627      & 7.17E-03 \\
$^{239}$U  & 8.78    & 87.6  & 0.742      & 8.47E-03 \\
$^{240}$U  & 2.67    & 24.3  & 2.33       & 9.59E-02 \\
$^{241}$U  & 8.02    & 36.8  & 1.88       & 5.11E-02 \\
$^{236}$Np & 0.171   & 40    & 0.0218     & 5.45E-04 \\
$^{236}$Np & 0.705   & 40    & 0.0432     & 1.08E-03 \\
$^{236}$Np & 2.02    & 40    & 0.152      & 3.80E-03 \\
$^{236}$Np & 2.41    & 40    & 0.0553     & 1.38E-03 \\
$^{236}$Np & 2.7     & 40    & 0.0971     & 2.43E-03 \\
$^{236}$Np & 2.77    & 40    & 0.071      & 1.78E-03 \\
$^{236}$Np & 3.13    & 40    & 0.0991     & 2.48E-03 \\
$^{236}$Np & 3.54    & 40    & 0.115      & 2.88E-03 \\
$^{236}$Np & 5.07    & 40    & 0.354      & 8.85E-03 \\
$^{236}$Np & 6.03    & 40    & 0.195      & 4.88E-03 \\
$^{236}$Np & 7.01    & 40    & 0.579      & 1.45E-02 \\
$^{236}$Np & 7.92    & 40    & 0.179      & 4.48E-03 \\
$^{236}$Np & 9.76    & 40    & 1.85       & 4.63E-02 \\
$^{237}$Np & 0.49    & 40.5  & 0.047      & 1.16E-03 \\
$^{237}$Np & 1.32    & 40.3  & 0.0323     & 8.01E-04 \\
$^{237}$Np & 1.48    & 40.5  & 0.184      & 4.54E-03 \\
$^{237}$Np & 1.97    & 39.5  & 0.0141     & 3.57E-04 \\
$^{237}$Np & 3.05    & 40.8  & 0.00000171 & 4.19E-08 \\
$^{237}$Np & 3.86    & 39.7  & 0.212      & 5.34E-03 \\
$^{237}$Np & 4.26    & 40.4  & 0.0326     & 8.07E-04 \\
$^{237}$Np & 4.86    & 40    & 0.0418     & 1.05E-03 \\
$^{237}$Np & 5.78    & 41.9  & 0.528      & 1.26E-02 \\
$^{237}$Np & 6.38    & 39.6  & 0.0789     & 1.99E-03 \\
$^{237}$Np & 6.68    & 40.1  & 0.0132     & 3.29E-04 \\
$^{237}$Np & 7.19    & 40    & 0.00888    & 2.22E-04 \\
$^{237}$Np & 7.42    & 38.4  & 0.122      & 3.18E-03 \\
$^{237}$Np & 7.68    & 40    & 0.00216    & 5.40E-05 \\
$^{237}$Np & 8.31    & 37.6  & 0.0902     & 2.40E-03 \\
$^{237}$Np & 8.98    & 37    & 0.102      & 2.76E-03 \\
$^{237}$Np & 9.3     & 41.4  & 0.602      & 1.45E-02 \\
$^{238}$Np & 0.181   & 50    & 0.118      & 2.36E-03 \\
$^{238}$Np & 1.12    & 50    & 0.308      & 6.16E-03 \\
$^{238}$Np & 1.83    & 50    & 0.149      & 2.98E-03 \\
$^{238}$Np & 2.29    & 50    & 0.0675     & 1.35E-03 \\
$^{238}$Np & 2.56    & 50    & 0.0621     & 1.24E-03 \\
$^{238}$Np & 2.88    & 50    & 0.205      & 4.10E-03 \\
$^{238}$Np & 3.28    & 50    & 0.132      & 2.64E-03 \\
$^{238}$Np & 3.53    & 50    & 0.134      & 2.68E-03 \\
$^{238}$Np & 4.04    & 50    & 0.0439     & 8.78E-04 \\
$^{238}$Np & 4.36    & 50    & 0.115      & 2.30E-03 \\
$^{238}$Np & 4.72    & 50    & 0.137      & 2.74E-03 \\
$^{238}$Np & 4.98    & 50    & 0.115      & 2.30E-03 \\
$^{238}$Np & 5.37    & 50    & 0.0388     & 7.76E-04 \\
$^{238}$Np & 5.75    & 50    & 0.0949     & 1.90E-03 \\
$^{238}$Np & 5.99    & 50    & 0.275      & 5.50E-03 \\
$^{238}$Np & 6.57    & 50    & 0.247      & 4.94E-03 \\
$^{236}$Pu & 3.16    & 44    & 1.56       & 3.55E-02 \\
$^{236}$Pu & 6.3     & 44    & 2.14       & 4.86E-02 \\
$^{238}$Pu & 2.89    & 38    & 0.0747     & 1.97E-03 \\
$^{238}$Pu & 9.98    & 37    & 0.208      & 5.62E-03 \\
$^{239}$Pu & 0.296   & 39.3  & 0.08       & 2.04E-03 \\
$^{239}$Pu & 7.82    & 37.7  & 0.792      & 2.10E-02 \\
$^{240}$Pu & 1.06    & 30    & 2.45       & 8.17E-02 \\
$^{241}$Pu & 0.15    & 42.3  & 0.000384   & 9.08E-06 \\
$^{241}$Pu & 0.264   & 34.6  & 0.0437     & 1.26E-03 \\
$^{241}$Pu & 1.73    & 40.3  & 0.00207    & 5.14E-05 \\
$^{241}$Pu & 4.29    & 34.7  & 0.576      & 1.66E-02 \\
$^{241}$Pu & 4.59    & 36.8  & 0.479      & 1.30E-02 \\
$^{241}$Pu & 5.81    & 60.1  & 2.77       & 4.61E-02 \\
$^{241}$Pu & 6.95    & 35.5  & 0.62       & 1.75E-02 \\
$^{241}$Pu & 8.62    & 33.6  & 0.78       & 2.32E-02 \\
$^{241}$Pu & 9.65    & 39.1  & 0.592      & 1.51E-02 \\
$^{241}$Pu & 9.94    & 56.6  & 1.89       & 3.34E-02 \\
$^{242}$Pu & 2.68    & 26.8  & 2          & 7.46E-02 \\
$^{243}$Pu & 1.66    & 73.4  & 0.556      & 7.57E-03 \\
$^{243}$Pu & 4.16    & 73.8  & 0.879      & 1.19E-02 \\
$^{243}$Pu & 6.66    & 74    & 1.11       & 1.50E-02 \\
$^{243}$Pu & 9.16    & 74.2  & 1.31       & 1.77E-02 \\
$^{244}$Pu & 4       & 20    & 0.026      & 1.30E-03 \\
$^{241}$Am & 0.305   & 44.4  & 0.079      & 1.78E-03 \\
$^{241}$Am & 0.572   & 43.3  & 0.134      & 3.09E-03 \\
$^{241}$Am & 1.27    & 45.3  & 0.325      & 7.17E-03 \\
$^{241}$Am & 1.92    & 41    & 0.144      & 3.51E-03 \\
$^{241}$Am & 2.36    & 50    & 0.0688     & 1.38E-03 \\
$^{241}$Am & 2.6     & 48    & 0.13       & 2.71E-03 \\
$^{241}$Am & 3.97    & 44.5  & 0.257      & 5.78E-03 \\
$^{241}$Am & 4.97    & 43.8  & 0.157      & 3.58E-03 \\
$^{241}$Am & 5.42    & 44.2  & 0.653      & 1.48E-02 \\
$^{241}$Am & 5.8     & 44.2  & 0.00247    & 5.59E-05 \\
$^{241}$Am & 6.12    & 43.8  & 0.112      & 2.56E-03 \\
$^{241}$Am & 6.75    & 44.2  & 0.0408     & 9.23E-04 \\
$^{241}$Am & 7.66    & 44.2  & 0.0423     & 9.57E-04 \\
$^{241}$Am & 8.17    & 42.7  & 0.0926     & 2.17E-03 \\
$^{241}$Am & 9.11    & 44.2  & 0.449      & 1.02E-02 \\
$^{241}$Am & 9.85    & 43.9  & 0.494      & 1.13E-02 \\
$^{242}$Am & 0.178   & 46    & 0.192      & 4.17E-03 \\
$^{242}$Am & 0.35    & 50    & 0.18       & 3.60E-03 \\
$^{242}$Am & 0.615   & 50    & 0.111      & 2.22E-03 \\
$^{242}$Am & 1.1     & 50    & 0.423      & 8.46E-03 \\
$^{242}$Am & 1.71    & 50    & 0.0504     & 1.01E-03 \\
$^{242}$Am & 2.11    & 50    & 0.181      & 3.62E-03 \\
$^{242}$Am & 2.19    & 50    & 0.286      & 5.72E-03 \\
$^{242}$Am & 2.31    & 50    & 0.152      & 3.04E-03 \\
$^{242}$Am & 2.95    & 50    & 0.0821     & 1.64E-03 \\
$^{242}$Am & 3.18    & 50    & 0.273      & 5.46E-03 \\
$^{242}$Am & 3.39    & 50    & 0.242      & 4.84E-03 \\
$^{242}$Am & 3.71    & 50    & 0.6        & 1.20E-02 \\
$^{242}$Am & 4.01    & 50    & 0.266      & 5.32E-03 \\
$^{242}$Am & 4.13    & 50    & 0.0019     & 3.80E-05 \\
$^{242}$Am & 4.27    & 50    & 0.234      & 4.68E-03 \\
$^{242}$Am & 4.55    & 50    & 0.231      & 4.62E-03 \\
$^{242}$Am & 5.37    & 50    & 0.526      & 1.05E-02 \\
$^{242}$Am & 5.7     & 50    & 0.0468     & 9.36E-04 \\
$^{242}$Am & 5.91    & 50    & 0.0806     & 1.61E-03 \\
$^{242}$Am & 5.95    & 50    & 0.356      & 7.12E-03 \\
$^{242}$Am & 6.15    & 50    & 0.0805     & 1.61E-03 \\
$^{242}$Am & 6.41    & 50    & 0.0128     & 2.56E-04 \\
$^{242}$Am & 6.65    & 50    & 0.214      & 4.28E-03 \\
$^{242}$Am & 6.71    & 50    & 0.13       & 2.60E-03 \\
$^{242}$Am & 6.84    & 50    & 0.038      & 7.60E-04 \\
$^{242}$Am & 6.99    & 50    & 1.74       & 3.48E-02 \\
$^{242}$Am & 7       & 50    & 0.0361     & 7.22E-04 \\
$^{242}$Am & 7.21    & 50    & 0.104      & 2.08E-03 \\
$^{242}$Am & 8.07    & 50    & 0.131      & 2.62E-03 \\
$^{242}$Am & 8.33    & 50    & 0.896      & 1.79E-02 \\
$^{242}$Am & 8.6     & 50    & 0.0733     & 1.47E-03 \\
$^{242}$Am & 8.86    & 50    & 0.00284    & 5.68E-05 \\
$^{242}$Am & 9.03    & 50    & 0.412      & 8.24E-03 \\
$^{242}$Am & 9.42    & 50    & 0.19       & 3.80E-03 \\
$^{242}$Am & 9.43    & 50    & 0.0569     & 1.14E-03 \\
$^{242}$Am & 9.88    & 50    & 0.159      & 3.18E-03 \\
$^{243}$Am & 0.419   & 22.9  & 0.000671   & 2.93E-05 \\
$^{243}$Am & 0.983   & 45    & 0.02       & 4.44E-04 \\
$^{243}$Am & 1.36    & 50    & 1.1        & 2.20E-02 \\
$^{243}$Am & 1.74    & 41.5  & 0.35       & 8.43E-03 \\
$^{243}$Am & 3.14    & 32    & 0.0136     & 4.25E-04 \\
$^{243}$Am & 3.42    & 38.7  & 0.235      & 6.07E-03 \\
$^{243}$Am & 3.85    & 44.9  & 0.0172     & 3.83E-04 \\
$^{243}$Am & 5.13    & 35.1  & 0.344      & 9.80E-03 \\
$^{243}$Am & 6.55    & 51.5  & 0.782      & 1.52E-02 \\
$^{243}$Am & 7.07    & 29.7  & 0.0545     & 1.84E-03 \\
$^{243}$Am & 7.86    & 46.9  & 1.48       & 3.16E-02 \\
$^{243}$Am & 8.38    & 38.4  & 0.00989    & 2.58E-04 \\
$^{243}$Am & 8.77    & 31.7  & 0.101      & 3.19E-03 \\
$^{243}$Am & 9.31    & 38.6  & 0.178      & 4.61E-03 \\
$^{243}$Cm & 0.671   & 40    & 0.0528     & 1.32E-03 \\
$^{243}$Cm & 1.14    & 40    & 0.0588     & 1.47E-03 \\
$^{243}$Cm & 1.47    & 40    & 0.0249     & 6.23E-04 \\
$^{243}$Cm & 2.05    & 40    & 0.0163     & 4.08E-04 \\
$^{243}$Cm & 2.31    & 40    & 1.93       & 4.83E-02 \\
$^{243}$Cm & 2.76    & 40    & 0.0137     & 3.43E-04 \\
$^{243}$Cm & 3.07    & 40    & 0.816      & 2.04E-02 \\
$^{243}$Cm & 3.73    & 40    & 0.54       & 1.35E-02 \\
$^{243}$Cm & 5.68    & 40    & 0.497      & 1.24E-02 \\
$^{243}$Cm & 6.15    & 40    & 1.73       & 4.33E-02 \\
$^{243}$Cm & 7.21    & 40    & 1.67       & 4.18E-02 \\
$^{243}$Cm & 8.18    & 40    & 0.394      & 9.85E-03 \\
$^{243}$Cm & 9.11    & 40    & 3.74       & 9.35E-02 \\
$^{244}$Cm & 7.67    & 37    & 9.6        & 2.59E-01 \\
$^{245}$Cm & 0.85    & 44    & 0.09       & 2.05E-03 \\
$^{245}$Cm & 1.98    & 44    & 0.2        & 4.55E-03 \\
$^{245}$Cm & 2.45    & 33    & 0.126      & 3.82E-03 \\
$^{245}$Cm & 4.68    & 33    & 1.87       & 5.67E-02 \\
$^{245}$Cm & 5.75    & 33    & 0.22       & 6.67E-03 \\
$^{245}$Cm & 7.53    & 33    & 2.8        & 8.48E-02 \\
$^{245}$Cm & 8.65    & 33    & 0.606      & 1.84E-02 \\
$^{245}$Cm & 9.15    & 33    & 0.347      & 1.05E-02 \\
$^{246}$Cm & 4.32    & 28    & 0.311      & 1.11E-02 \\
$^{247}$Cm & 1.24    & 32    & 0.665      & 2.08E-02 \\
$^{247}$Cm & 2.94    & 40    & 0.13       & 3.25E-03 \\
$^{247}$Cm & 3.18    & 40    & 1.06       & 2.65E-02 \\
$^{247}$Cm & 4.73    & 40    & 1.47       & 3.68E-02 \\
$^{247}$Cm & 6.12    & 40    & 0.105      & 2.63E-03 \\
$^{247}$Cm & 7.12    & 40    & 0.45       & 1.13E-02 \\
$^{247}$Cm & 7.65    & 40    & 0.115      & 2.88E-03 \\
$^{247}$Cm & 7.94    & 40    & 0.4        & 1.00E-02 \\
$^{247}$Cm & 9.5     & 40    & 0.72       & 1.80E-02 \\
$^{247}$Cm & 10      & 40    & 0.15       & 3.75E-03 \\
$^{248}$Cm & 7.25    & 23.3  & 1.89       & 8.11E-02 \\
$^{250}$Cm & 9.02    & 99.3  & 79.3       & 7.99E-01 \\
$^{249}$Bk & 0.195   & 35.9  & 0.117      & 3.26E-03 \\
$^{249}$Bk & 1.34    & 35.1  & 0.198      & 5.64E-03 \\
$^{249}$Bk & 1.6     & 33.2  & 0.573      & 1.73E-02 \\
$^{249}$Bk & 2.15    & 36.7  & 0.107      & 2.92E-03 \\
$^{249}$Bk & 3.11    & 37    & 0.145      & 3.92E-03 \\
$^{249}$Bk & 5.02    & 44.3  & 0.231      & 5.21E-03 \\
$^{249}$Bk & 6.28    & 33.8  & 0.147      & 4.35E-03 \\
$^{249}$Bk & 7.04    & 39    & 0.165      & 4.23E-03 \\
$^{249}$Bk & 7.99    & 36.1  & 1.41       & 3.91E-02 \\
$^{249}$Cf & 0.7     & 40    & 0.65       & 1.63E-02 \\
$^{249}$Cf & 3.88    & 40    & 0.291      & 7.28E-03 \\
$^{249}$Cf & 5.07    & 40    & 0.653      & 1.63E-02 \\
$^{249}$Cf & 7.51    & 40    & 0.147      & 3.68E-03 \\
$^{249}$Cf & 8.65    & 40    & 0.444      & 1.11E-02 \\
$^{249}$Cf & 9.51    & 40    & 1.61       & 4.03E-02 \\
$^{250}$Cf & 0.553   & 36.4  & 1.03       & 2.83E-02 \\
$^{250}$Cf & 6.85    & 45    & 0.425      & 9.44E-03 \\
$^{250}$Cf & 8.26    & 43    & 7.15       & 1.66E-01 \\
$^{251}$Cf & 0.389   & 35    & 0.877      & 2.51E-02 \\
$^{252}$Cf & 1.4     & 35    & 0.0192     & 5.49E-04 \\
$^{253}$Es & 0.7     & 24.4  & 0.65       & 2.66E-02 \\

\end{longtable*}

\end{document}